\documentclass{article}

\makeatletter
\renewcommand{\maketitle}{
  \begingroup
    \renewcommand{\thefootnote}{\fnsymbol{footnote}}
    \if@twocolumn
      \@topnewpage[\@maketitle]
    \else
      \newpage
      \global\@topnum\z@
      \@maketitle
    \fi
    \thispagestyle{empty}
    \@thanks
  \endgroup
  \setcounter{footnote}{0}%
}

\renewcommand{\@maketitle}{
  \newpage
  \null
  \vskip 2em
  \begin{center}
    {\LARGE \@title \par}
    \vskip 1em
    {\large \lineskip .5em
      \begin{tabular}[t]{c}
        \@author
      \end{tabular}\par}
    \vskip 1em
    {\large \@date}
  \end{center}
  \vskip 2em
}
\makeatother

\usepackage{pdflscape}
\usepackage[utf8]{inputenc}
\usepackage[T1]{fontenc} 
\usepackage{xcolor}
\definecolor{darkblue}{RGB}{0, 0, 100}

\usepackage{hyperref}
\hypersetup{
    colorlinks=true, 
    linkcolor=darkblue,
    citecolor=darkblue,
    urlcolor=darkblue
}

\usepackage{arxiv}
\usepackage{url} 
\usepackage{booktabs} 
\usepackage{amsfonts} 
\usepackage{nicefrac} 
\usepackage{amsmath} 
\usepackage{microtype} 
\usepackage{cleveref}  
\usepackage{graphicx}
\usepackage{natbib}
\usepackage{doi}
\usepackage{threeparttable}

\usepackage{amssymb}
\usepackage{amsthm}

\usepackage{subcaption}

\usepackage{setspace}
\usepackage{tabularx}

\usepackage{fancyhdr}
\usepackage[explicit]{titlesec}
\titleformat{\section}{\large\bfseries}{\thesection}{1em}{#1}

\usepackage{titling}
\usepackage{mathpazo}

\pretitle{\begin{center}\LARGE\bfseries}
\posttitle{\par\end{center}\vskip 0.5em}
\preauthor{\begin{center}\large}
\postauthor{\end{center}}
\predate{\begin{center}\small}
\postdate{\end{center}}

\begin{document}

\title{Distracting from the Epstein files?\\ Media attention and short-run shifts in Trump's Truth Social posts}

\author{\href{https://orcid.org/0000-0002-0811-3515}{\includegraphics[scale=0.06]{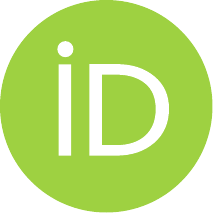}} \hspace{1mm} Andrew J. Peterson
\thanks{Assistant Professor (Ma\^{i}tre de conf\'{e}rences), University of Poitiers.}}

\renewcommand{\shorttitle}{Distracting from the Epstein files? }

\hypersetup{
pdftitle={Distracting from the Epstein files?},
pdfsubject={public opinion, accountability},
pdfauthor={Andrew J.~Peterson},
pdfkeywords={public opinion, accountability},
}

\date{\today}

\maketitle

\bigskip

\begin{abstract}
Political ``circuses'' may undermine democratic accountability if leaders facing scandal can reliably pull media coverage toward fresh topics and away from substantive investigations or evaluations. 
We investigate whether politicians strategically alter their messaging during damaging media coverage (``strategic diversion'') or maintain consistent provocative communication regardless of scandal coverage (``always-on circus''). Using computational text analysis of Donald Trump's Truth Social posts during the 2025 Epstein revelations,
we find that a one-standard-deviation increase in scandal coverage is associated with communication patterns that deviate from baseline by 0.28 standard deviations over a 4-day window. 
Although these findings do not provide formal causal identification, they are robust to timing placebos and falsification tests. They are consistent with the interpretation that leaders deploy diversionary communication specifically within their own friendly media ecosystem, with implications for accountability in polarized democracies.

\end{abstract}


\section{Introduction}
\label{sec:introduction}

A central tenet of democratic accountability is that the public, via the media, can evaluate and sanction leaders for their actions. 
This mechanism fails, however, if leaders can consistently evade scrutiny by strategically diverting attention to other issues. 
The dynamics of social media, which traditional media now frequently echo and amplify, offer a powerful tool for such agenda control. 

This study probes the limits of this power by examining whether a leader's strategic communication becomes \emph{more novel} when a persistent, damaging scandal flares. We treat linguistic novelty as an observable behavioral proxy that is consistent with attempts to redirect attention, but we do not attempt to measure leader mental states or intent, nor whether these attempts are ultimately successful.

We examine these dynamics in the context of the 2025 Epstein files scandal, a case that provides a clear example of a president facing a persistent, high-salience threat from both political opponents and internal party dissent. 
This case adds to existing work on diversionary communication in three ways. 
First, whereas much prior quantitative work has focused on legal or institutional threats such as investigations and corruption probes, we provide a detailed case study of a personal scandal that generated both reputational and intra-party pressure. 
Second, we measure the \emph{novelty} of the leader's content with an embeddings-based distance measure and extend the focus from Twitter and Facebook \citep[e.g.][]{barbera2024distract, lewandowsky2020tweets} to Truth Social, a smaller but more ideologically homogeneous platform with different dynamics. 
Third, by comparing responses to Epstein coverage in friendly and less-aligned media ecosystems, we use source-specific exposure to probe whether the president appears more responsive when the scandal surfaces in information environments central to his base.

We find that when Fox News attention to Epstein rises by one standard deviation, the short-run novelty of the president's posts increases by $0.28$ standard deviations over days $t$ through $t{+}3$. The response peaks within 2-3 days and persists for roughly a week in cumulative windows. Timing placebos (future exposure) are null, pre-trend diagnostics show no anticipatory pattern, and falsifications with unrelated media topics are null. The effect is strong for Fox and Google Trends, but not statistically significant for CNN and MSNBC. This aligns with the interpretation that Trump responds strategically to the media ecosystem he and his base consume most. We note that the observational nature of the study means we cannot definitively establish a causal relationship. We cannot rule out the possibility that an unobserved confounding factor (e.g., separate concurrent events) might be simultaneously influencing both the media's focus and the president's posting behavior.

The sections that follow explain narratively the case study context and our hypotheses (Section~\ref{sec:context}) and review related work (Section~\ref{sec:litreview}). We then detail the data and design (Section~\ref{sec:methods}), report the estimates and dynamics (Section~\ref{sec:results}), and the final section concludes with suggestions for additional work.  Additional details and robustness and specificity checks are in the Appendix, and replication code is available on \href{https://github.com/aristotle-tek/distract-epstein-replication}{Github}.

\section{The Epstein Scandal: Research Context and Hypotheses}
\label{sec:context}

\begin{figure}[!htbp]
  \centering
  \includegraphics[width=0.99\linewidth]{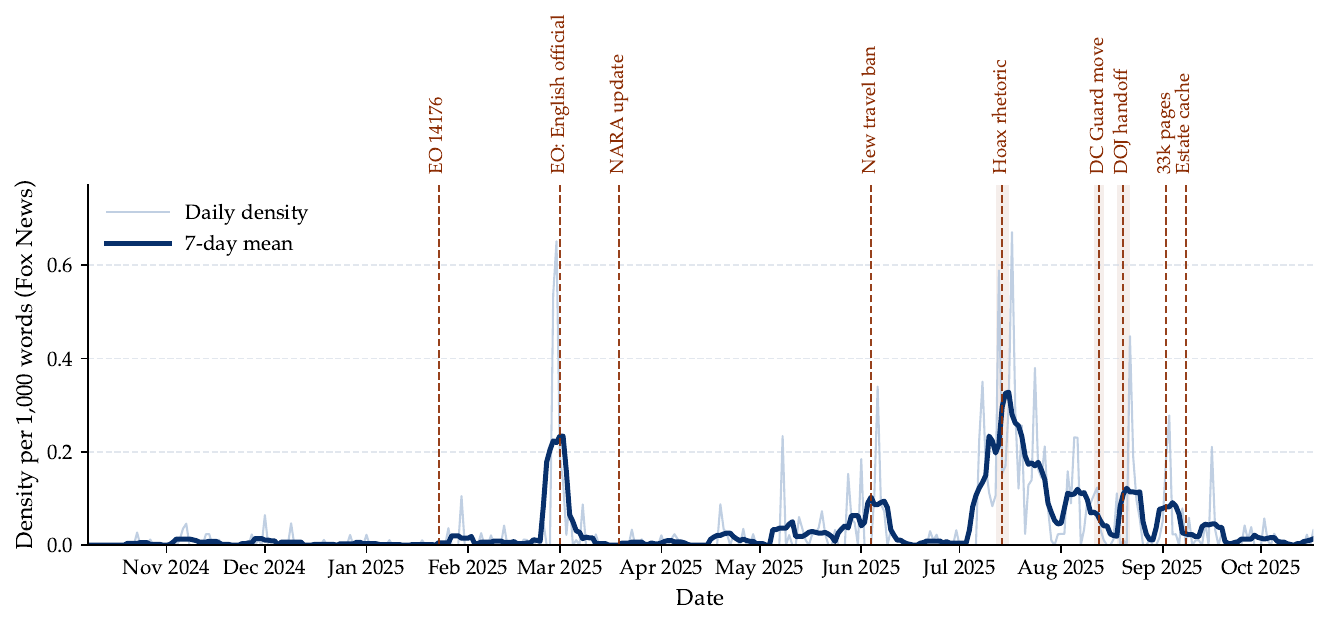}
  \caption{Density of appearances of `Epstein' in Fox News TV transcripts. Sources for events are provided in Appendix Table~\ref{tab:key_dates_sources}}
  \label{fig:timeline_attention}
\end{figure}

Since the beginning of 2025, the presidency faced recurring, high-salience public pressure to release the Epstein files (see Figure~\ref{fig:timeline_attention}). 
In response, the White House advanced a sequence of highly visible public countermoves, from large-scale declassifications of unrelated historical records (e.g., JFK/RFK/MLK files) to new law-and-order initiatives (e.g., deploying federal troops in Washington, D.C. in August)\footnote{See Appendix Table~\ref{tab:key_dates_sources} for dates and outlets.}, to announcing a \$10 billion lawsuit against the Wall Street Journal for publishing information about a 2003 birthday card for Epstein said to be written by Trump.\footnote{Bose \& Stempel, ``Trump sues Wall Street Journal over Epstein report, seeks \$10 billion'', July 19, 2025. \href{https://www.reuters.com/world/us/trump-sues-wall-street-journal-over-epstein-report-seeks-10-billion-2025-07-19/}{Reuters}.} These moves were coupled with a stream of attention-drawing social media posts, such as accusing former President Obama of ``treason'' (amplified by an AI-generated video of his fake arrest), alleging loan fraud against Adam Schiff, and musing about revoking Rosie O'Donnell's U.S. citizenship.
 
The political stakes of this scandal were high, as the threat originated not only from political opponents but also from within the president's own party. 
A populist faction of Republicans, many of whom have embraced conspiracy claims about Epstein’s death being a high-level cover-up, broke with the White House to demand full disclosure. This intra-party rebellion threatened to pass a discharge petition in the House for a bill to release the Epstein files.\footnote{See \href{https://clerk.house.gov/DischargePetition/2025090209}{discharge petition No. 9 (clerk.house.gov)} and \href{https://www.congress.gov/bill/119th-congress/house-resolution/581}{H.Res.581 (congress.gov)} }. The administration strongly sought to discourage this dissent, warning Republicans that supporting the petition would be viewed as a ``very hostile act to the administration''\footnote{Michael Gold, ``G.O.P. Thwarts Epstein Disclosure Bill as Accusers Plead for Files'', \href{https://www.nytimes.com/2025/09/03/us/politics/epstein-bill-republicans-trump.html}{New York Times} Sept 3, 2025. }.

This combined internal and external pressure created a recurring threat that the president was forced to manage, and was widely reported  as the reason why House Speaker Mike Johnson ``called the House to an early recess until September in order to avoid a floor vote on the [Epstein] legislation''\footnote{\href{https://www.foxnews.com/media/reps-massie-khanna-sound-off-whether-ghislaine-maxwell-should-receive-pardon-blast-mike-johnson}{ David Spector, Fox News,  July 27, 2025}} 
and then kept the House out of session to avoid seating a new House member who promised to sign the discharge petition, thereby tipping the vote.\footnote{\href{https://thehill.com/homenews/house/5529055-republicans-grijalva-swearing-in-house/}{The Hill}, \href{https://www.washingtonexaminer.com/news/house/3850807/gallego-mike-johnson-swear-in-grijalva-epstein/}{The Washington Examiner}, \href{https://www.wsj.com/politics/policy/bipartisan-pair-say-they-will-force-house-vote-on-releasing-epstein-files-after-recess-27b8ffd3}{The Wall Street Journal}, \href{https://www.newsweek.com/johnson-accused-delaying-grijalva-swearing-in-avoid-epstein-vote-10812792}{Newsweek}. On Representative Green (R) getting `pushback' from the White House on discharge bill position, see Timotija, ``Greene says she got `a lot' of pushback from White House over Epstein discharge petition'', \href{https://thehill.com/homenews/house/5485249-greene-pushback-white-house-epstein-discharge-petition/}{The Hill}, Sept. 3, 2025. }

 At moments of elevated Epstein attention, President Trump dismissed demands for a ``client list'' as a partisan ``hoax,'' castigated critics, and introduced diverse fresh, high-salience issues.  Contemporary reporting and commentary repeatedly framed these episodes as diversionary tactics.\footnote{Appendix Table~\ref{tab:diversions} provides an illustrative list.}

Our empirical design links the public attention environment to the leader's output on the Truth Social platform. While previous work has looked at Twitter and Facebook, Truth Social is unique in the extent to which it serves as the primary, high-volume broadcast channel for the president, but to a much smaller and more ideologically homogeneous user base.\footnote{Regular news consumers on Truth Social are overwhelmingly Republican or lean-Republican, and a majority of the platform's users utilize it for news. See \href{https://www.pewresearch.org/short-reads/2022/11/18/key-facts-about-truth-social-as-donald-trump-runs-for-u-s-president-again/}{Pew Research Center}, ``Key facts about Truth Social'' (Nov 18, 2022) and ``Social Media and News Fact Sheet'' (\href{https://www.pewresearch.org/journalism/fact-sheet/social-media-and-news-fact-sheet/}{Pew}; Nov 1, 2025). The platform may also have differential access, as suggested by an initial 6-hour exclusivity window before posting on other social media. While this agreement was scheduled to lapse unless renewed in February 2025, recent reporting (e.g. Amatulli, \href{https://www.theguardian.com/us-news/2025/nov/05/trump-truth-social-posts}{The Guardian} Nov. 6, 2025),  confirms that Truth Social remained the president's primary platform for high-volume communication.} This unique structure implies a different diversionary mechanism than the broad public square of Twitter: rather than engaging a mass public directly, its primary function is likely to mobilize the president's base and, crucially, to seed narratives for off-platform amplification by journalists and allied media outlets.\footnote{Reporting has noted that while on-platform engagement is comparatively small, posts are often picked up and broadcast by mainstream and conservative news, which becomes the primary vector for reaching a mass audience. See, e.g., \href{https://www.niemanlab.org/reading/are-people-paying-attention-to-trumps-truth-social-posts/}{Nieman Lab (2024)} and \href{https://www.bloomberg.com/news/articles/2024-12-03/trump-s-policy-posts-on-truth-social-fail-to-improve-traffic}{Bloomberg (2024)}.} This makes it an ideal setting to test whether the leader's \emph{novel} content (the supply) is designed to influence the broader media agenda (the measured outcome).

Investigative reporting (e.g., Mayer 2019 in \emph{\href{https://www.newyorker.com/magazine/2019/03/11/the-making-of-the-fox-news-white-house}{The New Yorker}}) has detailed how Fox personalities acted as informal advisers (including before being invited to join the administration). This pattern was observed in specific policy actions, such as directives on South Africa land reform or border troop deployments, which were announced in tweets minutes after corresponding Fox segments aired.\footnote{For quantitative evidence of this `watch-and-react' pattern, see \href{https://www.washingtonpost.com/politics/2018/12/31/times-that-trump-specifically-wanted-us-know-what-was-airing-fox-news/}{Philip Bump, ``The 135 times in 2018...'' (\emph{The Washington Post})}. On policy directives in response to coverage of South Africa land reform, see \href{https://www.washingtonpost.com/politics/2018/08/24/president-trumps-false-claim-about-murders-south-african-farms/}{\emph{The Washington Post}, 2018}) and on the deployment of troops to the border following coverage of a migrant caravan, see \href{https://abcnews.go.com/International/trump-warns-migrant-caravan-fox-news-report-organizers/story?id=54162057}{\emph{ABC News}, 2018}).} For this reason we specifically look at Fox News TV transcripts in contrast to other media sources as possible stimuli.

We consider two stylized interpretations of the president's communication style. The \emph{always-on circus} view treats provocative, attention-seeking communication as a standing feature of the modern presidency, with novelty driven primarily by a high but relatively stable baseline of controversy-seeking behavior. This perspective implies that, once we account for calendar effects and past posting patterns, day-to-day fluctuations in Epstein salience should not generate systematic shifts in novelty beyond that baseline. 
The diversionary hypothesis instead posits that when attention to Epstein rises, the president adjusts his messaging in an effort to dilute, redirect, or crowd out the damaging topic. Empirically, we test a necessary implication of the diversionary view: that, holding constant past novelty and calendar patterns, increases in Epstein salience are followed by systematic increases in novelty.\footnote{Evidence on this necessary condition cannot, on its own, rule out more moderate versions of the always-on circus idea,  but it does speak to whether the data are consistent with a fully static, scandal-insensitive baseline.}


\section{Related Literature}
\label{sec:litreview}

Our work is situated at the intersection of literature on strategic communication, media effects, and the institutional pressures of the modern presidency. The theoretical tension between diversion and a baseline political ``circus'' has origins that go far beyond the modern media era. 
The concept of diversion was famously captured in Juvenal's critique of Roman \textit{panem et circenses} (bread and circuses) 
\citep{juvenalX}. Machiavelli offered a more instrumental analysis, advising in \textit{The Prince} that rulers use spectacle as a calculated tool to manage reputation and project power, leaving the populace ``satisfied and stupefied'' \citep[][ch. 7]{machiavelli1532prince}. And a long tradition continued to analyze this as a method of control, and probe the nature of its relationship to public collective responses and the `crowd'.\footnote{While Thomas Hobbes, writing in \textit{Leviathan}, justified absolute sovereign control over public discourse as a necessity for maintaining peace rather than a diversionary tactic \citep{hobbes1651leviathan}, his work reinforces the importance of centralized narrative control. The psychological dynamics underlying apparently irrational reactions by crowds were also a focus of the late 19th century \citep{lebon1895psychologie,sighele2018criminal} and later picked up by \citet{freud1989group} and others.}

The contemporary theoretical foundation for the diversionary view is anchored in the international relations literature on diversionary war, which argues that leaders facing domestic turmoil may initiate external conflicts to divert public attention and rally support \citep{levy1989diversionary, tir2010territorial}. In the digital era, social media provides a low-cost, high-speed mechanism for such tactics, as \cite{barbera2024distract} demonstrate on a global scale. The most direct empirical precedent for our study is \citet{lewandowsky2020tweets}, who provide strong quantitative evidence that President Trump strategically used Twitter to distract from a political-legal threat. They find that spikes in media coverage of the Mueller investigation were followed by an increase in Trump's tweets on preferred topics (e.g., jobs, China, immigration), which in turn was followed by a measurable reduction in media coverage of the investigation. This demonstrates a clear, measurable instance of digital diversion.

In contrast, the always-on circus view is supported by institutional theories of the presidency. The concept of the ``permanent campaign'' suggests that the distinction between governing and campaigning has effectively dissolved, requiring leaders to engage in a continuous quest for public approval \citep{ornstein2000permanent}. 
This institutional pressure, which builds upon the 20th-century development of the ``rhetorical presidency'' and its norm of ``going public'' to legislate \citep{tulis2017rhetorical}, incentivizes a constant stream of provocative, attention-seeking communication. From this perspective, a political ``circus'' is the institutional constant rather than a reactive strategy. 

While the mechanisms of diversion were established in the foundational work on agenda-setting \citep{mccombs1977agenda} and priming \citep{iyengar2009news}, which explain how shifts in media salience alter public priorities and evaluative criteria, direct tests of digital diversion have only begun to explore how these dynamics are evolving. We extend this line of research by analyzing a personal scandal, leveraging variation between friendly and mainstream outlets, and focusing on the novelty and content of the president's Truth Social posts themselves, thereby speaking to both diversionary theory and accounts of the modern political spectacle \citep{edelman1988constructing}.


\section{Methods}
\label{sec:methods}

\subsection{Data and Preprocessing}
\label{subsec:data}

We construct a daily time–series to estimate the dynamic relationship between media attention and content novelty. The estimation sample, after trimming for lags/leads, spans October 8, 2024–October 17, 2025.\footnote{The Trump Truth Social data comes from a \href{https://github.com/stiles/trump-truth-social-archive}{Github repository}, which is no longer collecting data}. In the results we present, we restrict the sample to 
$t$ has $\geq 3$ posts and the trailing window contributes $\geq 10$ posts (see Appendix Table~\ref{tab:post_intensity} for more detail), resulting in $n=256$ observations.

The dependent variable $Y_t$ is the z-scored novelty index $N_{t,z}$. In our baseline, novelty is measured by energy distance on whitened embeddings, with robustness to an alternative MMD$^2$ measure (Section~\ref{app:novelty}).\footnote{By construction, $N_t$ increases whenever the joint distribution of embeddings on day $t$ is unusual relative to the recent reference window, regardless of whether posts mention Epstein or not. High novelty can therefore arise from (i) shifts toward entirely different issues, (ii) new framings or rhetoric around the same issue or (iii) changes in style or emphasis that are not easily described in simple topic labels. Our empirical analyses below speak to whether this behavioral signal of novelty co-moves with Epstein salience; they do not, on their own, identify which of these substantive channels is operating.}

The main independent variable $E_t$ is ``Epstein attention.'' Our baseline uses per-1,000-word counts from Fox News TV transcripts from the Internet Archive;\footnote{\href{https://archive.org/details/tvarchive}{Internet Archive Television Archive}} robustness compares CNN/MSNBC transcripts and Google Trends (Appendix Table~\ref{tab:appendix_exposure}), in line with work using Google search as an attention proxy \citep{da2011attention, mellon2014google}. Exposure is standardized to $E_{t,z}$. Controls $X_t$ include day-of-week and month fixed effects, a post-inauguration indicator, and same-day posting-intensity measures. Availability windows differ slightly across exposure series.\footnote{For example, Google Trends coverage begins on 2024-11-04, yielding smaller $n$ in those specifications (see Appendix Table \ref{tab:appendix_exposure}).}

\subsection{Estimation and Inference}

We estimate autoregressive distributed-lag (ARDL) regressions
\begin{equation}
\label{eq:ardl}
Y_t = \alpha + \sum_{i=1}^{p} \phi_i\, Y_{t-i} + \sum_{j=0}^{q} \beta_j\, E_{t,z-j} + \Gamma' X_t + \varepsilon_t,
\end{equation}
with $p=7$ and $q=3$. The short-run cumulative effect is
\begin{equation}
\label{eq:sum}
\beta_{\text{sum}} = \sum_{j=0}^{q} \beta_j.
\end{equation}

To probe timing and potential reverse causality, we augment equation~\eqref{eq:ardl} with $L = 3$ leads of exposure and test whether the cumulative effect of these leads is zero:
\begin{equation}
\label{eq:leads}
Y_t = \alpha + \sum_{i=1}^{p} \phi_i\, Y_{t-i} + \sum_{j=0}^{q} \beta_j\, E_{t,z-j} + \sum_{h=1}^{L} \delta_h\, E_{t,z+h} + \Gamma' X_t + \varepsilon_t,
\end{equation}
with $\delta_{\text{sum}} = \sum_{h=1}^{L} \delta_h$.

All ARDL models are estimated by OLS with Newey–West heteroskedasticity- and autocorrelation-consistent standard errors \citep{newey1987simple}, using a weekly bandwidth ($H = 7$). For $\beta_{\text{sum}}$ and $\delta_{\text{sum}}$ we report parametric Wald tests based on the HAC covariance matrix. Unless otherwise noted, all reported $p$-values in the main text and appendix are based on these HAC standard errors.

\subsection{Impulse-Response Analyses}

To visualize the dynamic response of novelty to exposure shocks, we estimate impulse responses using local projections \citep{jorda2005local}. For each horizon $h$, we run
\[
Y_{t+h} = \alpha_h + \theta_h E_{t,z} + \sum_{i=1}^{p} \phi_{h,i}\, Y_{t-i} + \Gamma_h' X_t + u_{t+h},
\]
with $p = 7$ lags of $Y_t$, the same calendar controls as in the ARDL models, and Newey–West HAC($H = 7$) standard errors. We plot the estimated response coefficients $\theta_h$ along with conventional pointwise HAC confidence intervals.

Joint pre-trend checks over negative horizons use Wald tests of the null that the pre-exposure coefficients are jointly zero. We report HAC-based Wald $p$-values for these pre-trend tests (Appendix Table~\ref{tab:appendix_pretrend}).


\section{Results}
\label{sec:results}

Our empirical analysis documents a clear, statistically significant, and substantively meaningful association between increases in media attention to the Epstein files and subsequent shifts in the novelty of Donald Trump's Truth Social communications. 
The timing, specificity, and source-dependence of this relationship are more consistent with a departure from a purely static, “always-on circus” view than with it, and are suggestive of strategic communications responsiveness to the scandal. 
At the same time, the observational, single-case design means these patterns should be interpreted as suggestive rather than definitive evidence of diversion.\footnote{We do not attempt to distinguish whether Trump changes topics specifically or simply varies his rhetorical style in a way captured by embedding distance while remaining on the same broad topics}.

\subsection{Main ARDL Estimates}

Table \ref{tab:main_ardl} presents the core findings from our ARDL models. 
Our primary specification ($q=3$), which models content novelty as a function of contemporaneous and three lagged days of Epstein exposure, yields a cumulative short-run effect ($\widehat{\beta}_{\text{sum}}$) of $0.285$. 
This estimate, which holds after accounting for day-of-week and monthly fixed effects, is significant under conventional HAC inference ($p < 0.001$) and is stable across reasonable variations in lag length and control sets (Appendix Table~\ref{tab:appendix_spec_grid}).

\begin{table}
\centering

\caption{ARDL short-run effects of Epstein exposure on novelty.}
\label{tab:main_ardl}
\begin{tabular}{lrrrrr}
\toprule
Spec & $\widehat{\beta}_{\text{sum}}$ & HAC s.e. & $p_{\text{HAC}}$ & $n$ & $R^2$ \\
\midrule
$q=1$ & 0.104 & 0.061 & 0.090 & 256 & 0.198 \\
$q=3$ & 0.285 & 0.077 & <0.001 & 256 & 0.226 \\
$q=7$ & 0.481 & 0.124 & <0.001 & 256 & 0.251 \\
Leads sum ($L=3$) & -0.042 & 0.071 & 0.553 & 255 & 0.234 \\
\bottomrule
\end{tabular}
\caption*{Entries are sums of exposure-lag coefficients with heteroskedasticity-and-autocorrelation-consistent (HAC) standard errors. The final row reports the sum of exposure leads as a timing placebo.}
\end{table}

\bigskip

\begin{figure}[!htbp]
  \centering
  \includegraphics[width=0.5\linewidth]{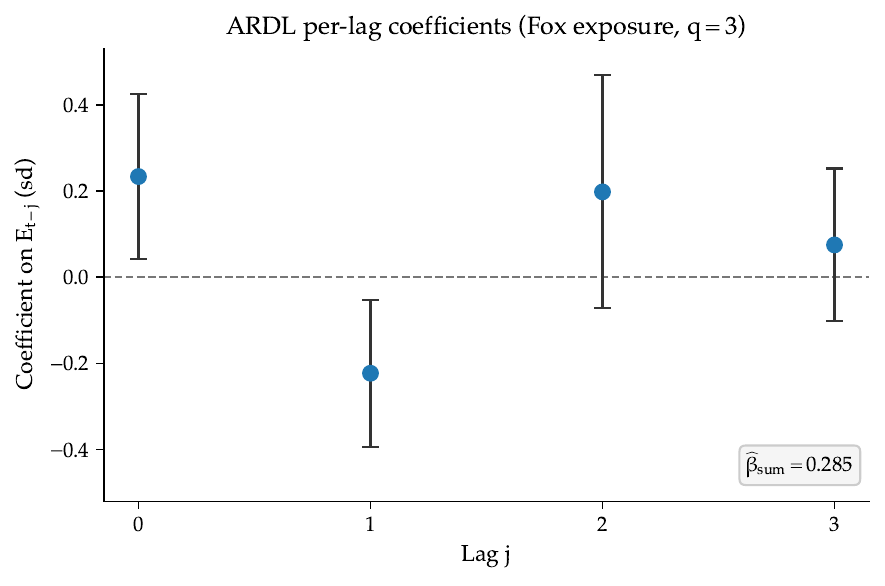}
  \caption{Per-lag ARDL coefficients ($q{=}3$) with 95\% HAC CIs. Box annotates the short-run sum $\widehat\beta_{\text{sum}}$.}
  \label{fig:perlag}
\end{figure}

In substantive terms, this coefficient implies that a one standard deviation increase in Fox News attention to the Epstein story is associated with a cumulative 0.28 standard deviation increase in the novelty of Trump's posts over the four-day window (lags 0–3). The effect is not only immediate but also accumulates over several days, as illustrated by the per-lag coefficients in Figure \ref{fig:perlag}. 
In short: novelty jumps the same day Fox attention rises, but reverses the next day, then stays modestly elevated, adding up to a meaningful short-run effect. Overall, this provides clear evidence against the idea that there is no increase in novelty after Epstein attention spikes.

\subsection{Timing, Dynamics, and Threats to Inference}

A central challenge in political communication research is distinguishing strategic action from mere correlation. Our research design incorporates two diagnostic tests that help assess whether the observed association is consistent with a causal interpretation or more likely reflects simple reverse causality or anticipation.

First, we employ a ``timing placebo'' test by including future (lead) values of the exposure variable in our model. If the relationship were spurious, or if Trump's novel posting \emph{caused} the media attention, we might expect to see a relationship with future media hits. 
The final row of Table \ref{tab:main_ardl} shows this is not the case. The cumulative sum of the lead coefficients is small, negative, and statistically indistinguishable from zero ($\widehat{\delta}_{\text{sum}} = -0.042$, $p = 0.553$). 
This null finding demonstrates that future increases in Epstein attention do not predict current shifts in post novelty, ruling out simple anticipation effects or reverse causality.

\begin{figure}[!htbp]
  \centering
  \includegraphics[width=0.8\linewidth]{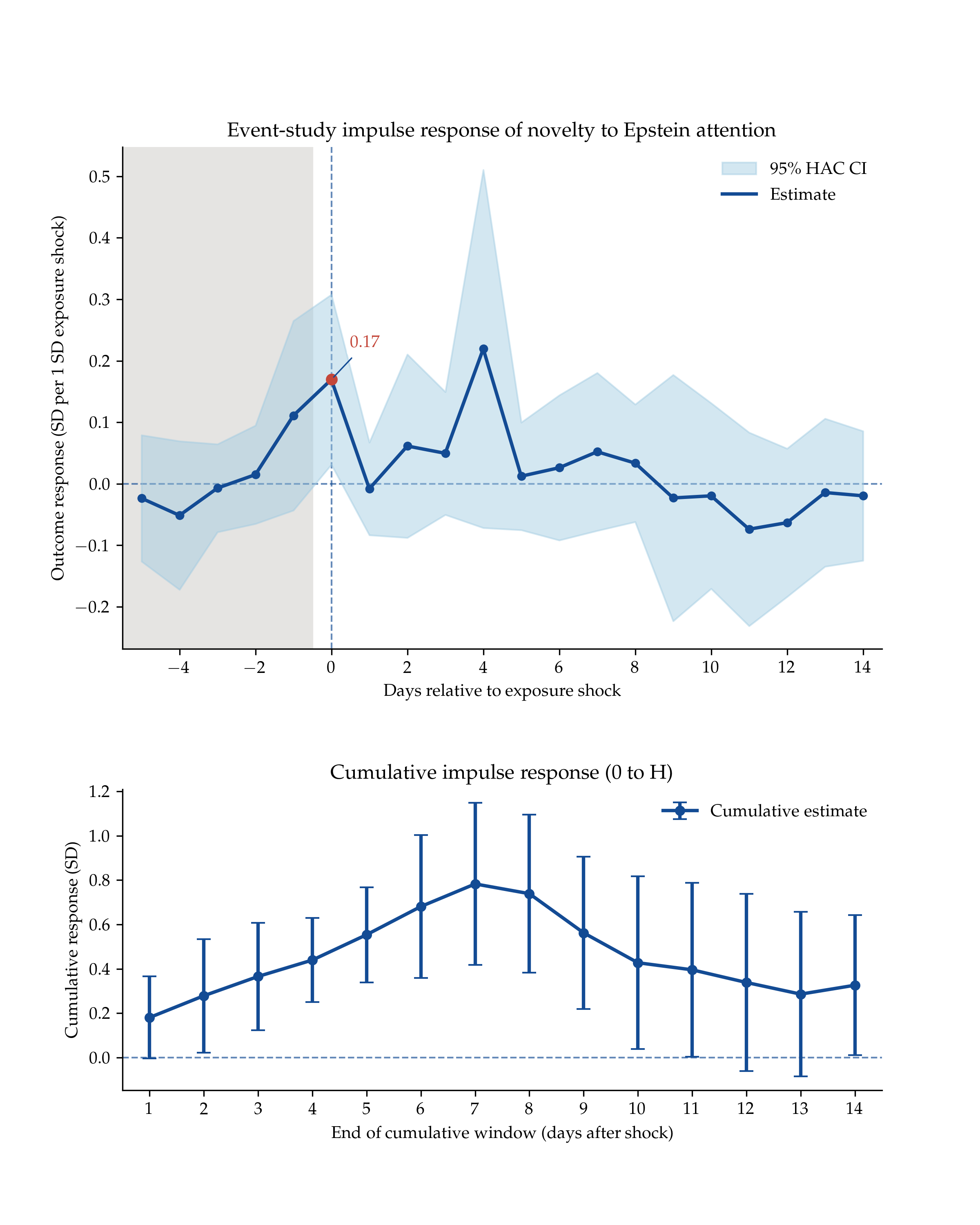}
  \caption{Local-projection estimates of $Y_{t+h}$ on $E_t$ with $p{=}7$ $Y$-lags and calendar FE. Top: level IRF; bottom: cumulative IRF (sum of responses through horizon $h$). Shaded bands show 95\% HAC CIs. Variables are standardized; coefficients read as sd($Y$) per 1 sd($E$). Vertical line marks $h{=}0$; $h{<}0$ are pre-trend placebos.}
  \label{fig:irf}
\end{figure}

Second, the impulse response functions (IRFs) estimated via local projections (Figure \ref{fig:irf}) visualize this dynamic relationship. The plot confirms the findings from the leads test: there is no pre-trend, with coefficients for horizons $h < 0$ tightly clustered around zero (see also Table \ref{tab:appendix_pretrend} for formal tests). 
The response to an increase in attention is, however, immediate. 
The effect begins at $h=0$, peaks within 2–3 days, and remains statistically significant for approximately one week. 
This temporal pattern—a sharp, immediate, and persistent-but-decaying response—is consistent with what one would expect from a strategic actor reacting in real time to a negative attention spike, although we cannot rule out all other explanations.

\subsection{Robustness and specificity of the effect}

The observed effect is not an artifact of our measurement strategy, nor is it a generic response to any media ``buzz.'' 
We test for topical specificity using a falsification test. In Table \ref{tab:falsification_tv}, we replace our Epstein-related exposure variable with other high-salience, but politically unrelated, media topics: ``Taylor Swift'' and ``NCAA basketball.'' Both placebo exposures yield small and statistically weak effects; HAC $p$-values are $0.098$ for ``Taylor Swift'' and comfortably above conventional thresholds for ``NCAA basketball.'' This pattern implies the observed response is not a generic reaction to media excitement, but rather a targeted response specific to the politically damaging Epstein coverage.

We compute $N_t$ only on adequately sampled days ($\geq 3$ posts and $\geq 10$ trailing-reference posts) and include posting-intensity controls in all ARDLs, and diagnostics show $N_t$ is effectively uncorrelated with volume ($r = -0.054$) with 6.5\% low-sample days and 1.0\% zero-post days; see Table~\ref{tab:post_intensity} for a compact summary and Table~\ref{tab:top_novelty} for top novelty dates.

\begin{figure}[!htbp]
  \centering
  \includegraphics[width=0.7\linewidth]{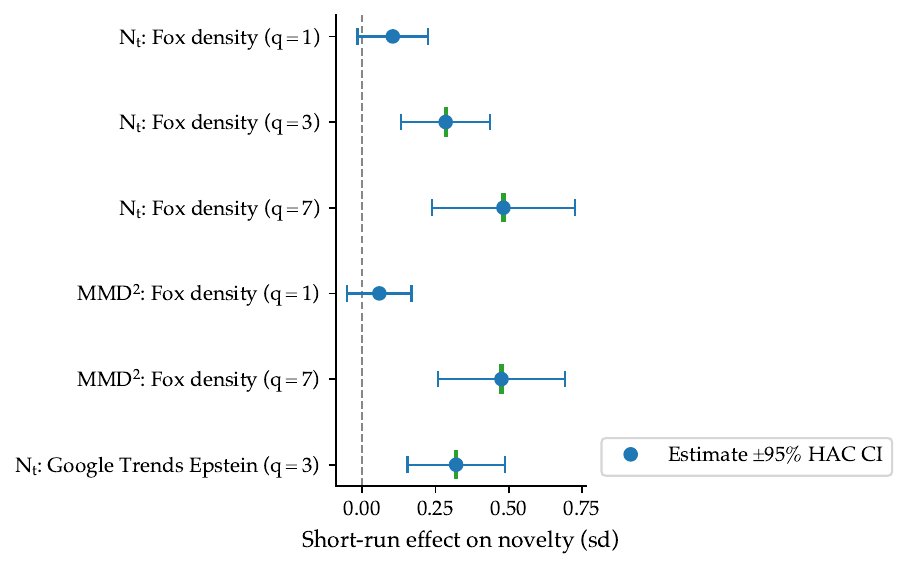}
  \caption{Robustness of short-run effect across exposure and novelty measures. Points show $\widehat\beta_{\text{sum}}$ with 95\% HAC CIs.}
  \label{fig:forest}
\end{figure}

The finding is robust to alternative measures of both the dependent and independent variables, as shown in the forest plot (Figure \ref{fig:forest}). The positive and significant effect holds when using an alternative novelty metric (Squared Maximum Mean Discrepancy, Table \ref{tab:mmd2_ardl}) and a different measure of public attention (Google Trends, Table \ref{tab:appendix_exposure}). 
Although the series show differing forms of stationary behavior, they are both effectively I(0), which poses no problem for an ARDL specification as long as standard deterministic terms are included.\footnote{In level tests, $N_{t,z}$ is clearly stationary (ADF $p \approx 0.000$, KPSS $p \approx 0.10$), while the z-scored Fox-Epstein exposure rejects a unit root (ADF $p \approx 0.005$) but fails KPSS level-stationarity ($p \approx 0.01$). In trend-augmented tests, $N_{t,z}$ still strongly rejects a unit root (ADF $p \approx 0.000$) but marginally fails KPSS trend-stationarity (KPSS $p \approx 0.03$), whereas the exposure rejects a unit root (ADF $p \approx 0.012$) and is consistent with trend-stationarity (KPSS $p \approx 0.10$). While the main results include an intercept but not a deterministic time trend, adding this linear time trend does not alter the main ARDL result. The short-run cumulative effect remains positive and statistically significant: $\beta_{\text{sum}} = 0.262$ (SE $= 0.076$; $p < 0.01$); $N = 256$; $R^2 = 0.241$. Inference is unchanged relative to the baseline specification.}

Finally, the analysis reveals a critical nuance in exposure source. The effect is strong and statistically precise for attention on Fox News and in Google Trends, but we do not detect comparable associations for coverage on CNN or MSNBC in this sample (Table~\ref{tab:appendix_exposure}). Taken at face value, this pattern is consistent with the idea that responsiveness is concentrated when the scandal surfaces in media environments that are especially salient for the president and his supporters (Fox News) or in broader public attention (Google Trends), rather than when it appears in less-aligned outlets. However, as the design does not directly test differences in coefficients across outlets, we interpret these source differences cautiously.

While these findings provide robust evidence for a diversionary response, some interpretative caution is warranted. Our methods establish a strong temporal relationship that is inconsistent with simple reverse causality, but like all observational time-series, cannot definitively rule out a complex, unobserved confounder that simultaneously drives both media attention and posting behavior.

Furthermore, our measure of novelty quantifies a shift in \emph{linguistic} patterns; whether this represents a deliberate pivot to new topics or simply a change in rhetorical style remains an important distinction. Finally, this study examines one actor, in response to one event, on one platform. While the finding that the response is triggered by friendly media (Fox) and public search (Google) provides a crucial clue, the precise human mechanism, be it a principal-level reaction or a staff-driven strategy, is beyond the scope of this data, and the external validity of these findings remains a question for future research. 
These limitations notwithstanding, the overall pattern is more consistent with strategic responsiveness than with a fully static, scandal-insensitive baseline.


\section{Conclusion}
\label{sec:conclusion}

We set out to explore whether a president's provocative communication style is a static, baseline feature (an ``always-on circus''), or is consistent with a strategic tool for managing political threats. Our analysis of the 2025 Epstein scandal is more consistent with the latter interpretation. 
We find that increases in Epstein attention on Fox News, or in Google search interest, are followed by rapid and statistically significant increases in the linguistic novelty of the president's Truth Social output, whereas we do not detect comparable patterns for coverage on CNN or MSNBC. A fruitful avenue for future work would be to examine the specific dynamics of which co-partisans willingly challenge a scandal-plagued but powerful leader, and how this interacts with the media environment. 

The response is not likely a generic reaction to general media shifts, as evidenced by our null falsification tests, nor is it simply reverse causality, as shown by our timing placebos. 
Instead, the evidence is consistent with targeted strategic responsiveness. In particular, the pattern of media-source heterogeneity fits the interpretation that the president is especially responsive when a damaging narrative surfaces in outlets that are central to his base's information environment, although the underlying decision-making process is beyond the scope of our data. Further work could specifically aim to identify the nature of how these two signals differ.

A critical limitation is that while we find the presence of a strategic response in linguistic novelty, we do not investigate its specific content. Our embedding-based measures show that the president's messaging becomes more unusual relative to his recent content, but they do not tell us whether this novelty comes from moving onto different topics, reframing, escalating attacks on opponents, or changing rhetorical style. As a result, novelty should be viewed as a behavioral signal that is consistent with, but not restricted to, diversionary communication. As an initial descriptive exercise, Figure~\ref{fig:truth_bucket_epstein_daily} provides a timeline that juxtaposes the daily density of Epstein mentions in Fox News TV transcripts with the density of Trump's Truth Social posts across selected topical categories. We also do not know to what extent these shifts influenced media uptake or ultimately shifted public opinion. 

Finally, we cannot definitively rule out all unobserved confounders and do not make a claim about causality or about the intent or mental states of a leader. The analysis is based on observational data for a single leader, in response to one particular scandal, on a single social media platform. As such, the results should be viewed as a detailed case study of strategic responsiveness rather than a general law of political communication. The external validity of these dynamics, such as whether they apply to other leaders, different platforms, or more conventional policy-related threats, remains an open question for future research.
Because both the scandal and the broader communications environment continue to evolve, the temporal and external validity of these specific estimates is inherently limited. On the other hand, the design and measurement choices we lay out here can be read as a kind of elaborate pre-registration that future work can revisit and extend as additional data become available.

These limitations provide a clear roadmap for future inquiry. The most immediate and vital next step is to open the black box of these novelty spikes. 
More broadly, this research agenda is central to understanding the mechanics of modern democratic accountability. This study provides behavioral evidence for the supply side of agenda control. The critical, unanswered question is to what degree it works, and when. Future research should test for downstream effects: do these spikes in communicative novelty successfully suppress media coverage of the original scandal or blunt its impact on public opinion? If leaders can, in fact, use the high-velocity attention economy to systematically crowd out damaging narratives, it poses a profound challenge to the public's ability to scrutinize and sanction their actions.


\bigskip



\bibliographystyle{plainnat}
\bigskip

\bibliography{diversions}

@article{newey1987simple,
  title={A Simple, Positive Semi-Definite, Heteroskedasticity and Autocorrelation Consistent Covariance Matrix},
  author={Newey, Whitney K. and West, Kenneth D.},
  journal={Econometrica},
  volume={55},
  number={3},
  pages={703--708},
  year={1987},
  publisher={Wiley}
}

@article{jorda2005local,
  title={Estimation and Inference of Impulse Responses by Local Projections},
  author={Jord{\`a}, {\`O}scar},
  journal={American Economic Review},
  volume={95},
  number={1},
  pages={161--182},
  year={2005},
  doi={10.1257/0002828053828518}
}

@article{szekely2007measuring,
  title={Measuring and testing dependence by correlation of distances},
  author={Sz{\'e}kely, GJ and Rizzo, ML and Bakirov, NK},
  journal={Annals of Statistics},
  volume={35},
  number={6},
  pages={2769--2794},
  year={2007},
  publisher={Institute of Mathematical Statistics}
}

@article{szekely2013energy,
  title={Energy statistics: A class of statistics based on distances},
  author={Sz{\'e}kely, G{\'a}bor J and Rizzo, Maria L},
  journal={Journal of statistical planning and inference},
  volume={143},
  number={8},
  pages={1249--1272},
  year={2013},
  publisher={Elsevier}
}

@article{gretton2012kernel,
  title={A kernel two-sample test},
  author={Gretton, Arthur and Borgwardt, Karsten M and Rasch, Malte J and Sch{\"o}lkopf, Bernhard and Smola, Alexander},
  journal={The journal of machine learning research},
  volume={13},
  number={1},
  pages={723--773},
  year={2012},
  publisher={JMLR. org}
}

@inproceedings{li2020sentence,
  title={On the Sentence Embeddings from Pre-trained Language Models},
  author={Li, Bohan and Zhou, Hao and He, Junxian and Wang, Mingxuan and Yang, Yiming and Li, Lei},
  booktitle = "Proceedings of the 2020 Conference on Empirical Methods in Natural Language Processing (EMNLP)",
  month = nov,
  pages={9119--9130},
  year={2020},
  doi = {10.18653/v1/2020.emnlp-main.733}
}

@article{su2021whitening,
  title={Whitening sentence representations for better semantics and faster retrieval},
  author={Su, Jianlin and Cao, Jiarun and Liu, Weijie and Ou, Yangyiwen},
  journal={arXiv preprint arXiv:2103.15316},
  year={2021}
}

@book{sighele2018criminal,
  title={The criminal crowd and other writings on mass society},
  author={Sighele, Scipio},
  year={1903},
  publisher={University of Toronto Press},
  note = {(L'intelligenza della folla). This edition 2018}
}

@book{lebon1895psychologie,
  title={Psychologie des foules},
  author={Le Bon, Gustave},
  year={1895},
  publisher={Alcan}
}

@book{freud1989group,
  title={Group psychology and the analysis of the ego},
  author={Freud, Sigmund},
  year={1921},
  publisher={WW Norton \& Company},
  note = {Originally published 1921, This edition 1989}
}

@book{edelman1988constructing,
  title={Constructing the political spectacle.},
  author={Edelman, Murray},
  year={1988},
  publisher={University of Chicago Press}
}

@article{lewandowsky2020tweets,
  title={Using the president's tweets to understand political diversion in the age of social media},
  author={Lewandowsky, Stephan and Jetter, Michael and Ecker, Ullrich KH},
  journal={Nature communications},
  volume={11},
  number={1},
  pages={5764},
  year={2020},
  publisher={Nature Publishing Group UK London}
}

@book{iyengar2009news,
  title={News that matters: Television \& American opinion},
  author={Iyengar, Shanto and Kinder, Donald R},
  year={2009},
  publisher={University of Chicago Press}
}

@article{barbera2024distract,
author = {Pablo Barberá and Anita R. Gohdes and Evgeniia Iakhnis and Thomas Zeitzoff},
title ={Distract and Divert: How World Leaders Use Social Media During Contentious Politics},
journal = {The International Journal of Press/Politics},
volume = {29},
number = {1},
pages = {47-73},
year = {2024},
doi = {10.1177/19401612221102030}
}

@article{levy1989diversionary,
  title={The diversionary theory of war: A critique},
  author={Levy, Jack S},
  journal={Handbook of war studies},
  volume={1},
  pages={259--288},
  year={1989}
}

@book{tulis2017rhetorical,
  title        = {The Rhetorical Presidency: New Edition},
  author       = {Tulis, Jeffrey K},
  year         = {2017},
  publisher    = {Princeton University Press}
}

@article{tir2010territorial,
  title={Territorial diversion: Diversionary theory of war and territorial conflict},
  author={Tir, Jaroslav},
  journal={The Journal of Politics},
  volume={72},
  number={2},
  pages={413--425},
  year={2010},
  publisher={Cambridge University Press New York, USA}
}

@article{mccombs1977agenda,
  title={Agenda setting function of mass media},
  author={McCombs, Maxwell},
  journal={Public relations review},
  volume={3},
  number={4},
  pages={89--95},
  year={1977},
  publisher={Elsevier}
}

@book{ornstein2000permanent,
  title={The permanent campaign and its future},
  author={Ornstein, Norman J},
  year={2000},
  publisher={American Enterprise Institute}
}

@misc{juvenalX,
  title = {The Satires {X}},
  author = {Juvenal},
  journall = {The Latin Library},
  url = {https://www.thelatinlibrary.com/juvenal/10.shtml},
  urldate = {2025-10-22},
  year      = {1893},
  note      = {From B{\"u}cheler's text of 1893; composed ca.\ 101 CE. Accessed 22 Oct 2025.}
}

@book{machiavelli1532prince,
  author       = {Niccol{\`o} Machiavelli},
  title        = {The Prince},
  year         = {1532},
  publisher    = {Penguin Classics},
  address      = {London},
  edition      = {Revised edition},
  translator   = {George Bull},
  note         = {Originally published in 1532 (written 1513), this edition 2003.}
}

@book{hobbes1651leviathan,
  author       = {Thomas Hobbes},
  title        = {Leviathan},
  year         = {1651},
  publisher    = {Penguin Classics},
  address      = {London},
  note         = {Edited by C. B. Macpherson. This edition 1982.}
}

@article{da2011attention,
  title={In Search of Attention},
  author={Da, Zhi and Engelberg, Joseph and Gao, Pengjie},
  journal={The Journal of Finance},
  volume={66},
  number={5},
  pages={1461--1499},
  year={2011},
  publisher={Wiley Online Library}
}

@article{mellon2014google,
  title={Internet Search Data and Issue Salience: The Properties of Google Trends as a Measure of Issue Salience},
  author={Mellon, Jonathan},
  journal={Journal of Elections, Public Opinion \& Parties},
  volume={24},
  number={1},
  pages={45--72},
  year={2014},
  publisher={Taylor \& Francis}
}

\appendix

\section{Appendix}
\label{sec:appendix}

\subsection{Embedding and Novelty Measurement}
\label{app:novelty}

To quantify content novelty, we first convert the text of each post into a high-dimensional vector representation (embedding). This process begins by cleaning the raw post content (e.g., stripping HTML tags) and feeding the text into a pre-trained \texttt{SentenceTransformer} model, specifically \texttt{all-MiniLM-L6-v2}. This model maps each post to a 384-dimensional vector. From the full corpus of $N$ posts, we obtain a matrix of ``raw" embeddings. 

These raw embeddings are known to suffer from anisotropy (a non-uniform distribution in the vector space), which can make distance metrics unreliable \citep{li2020sentence}. To correct this, we apply a standard decorrelation step. We fit a Principal Component Analysis model with whitening to the entire matrix $\mathbf{E}_{\text{raw}}$. This transformation de-correlates the features and scales them to have unit variance, yielding a matrix of `whitened' embeddings, $\mathbf{E}_{\text{white}}$ \citep{su2021whitening}. These whitened vectors are used for all novelty calculations.

The daily novelty measure, $N_t$, is constructed as a two-sample distributional distance. It quantifies how different the distribution of content on a given day $t$ is from the distribution of content in the recent past. We define two samples for each day $t$:
\begin{itemize}
    \item Sample 1 ($Y_t$): The set of whitened embeddings $\{\mathbf{e}_i \in \mathbf{E}_{\text{white}}\}$ for all posts published on day $t$.
    \item Sample 2 ($Y_{\text{ref}}$): The set of whitened embeddings for all posts published in a trailing reference window of $W$ days (e.g., days $t-W$ through $t-1$).
\end{itemize}
Prior to computing distances, all vectors in both $Y_t$ and $Y_{\text{ref}}$ are $L_2$-normalized (unit-normalized). The novelty score $N_t$ is therefore a measure of distributional shift on the unit hypersphere in the decorrelated PCA space.

Our primary novelty measure, used for the main results (Table \ref{tab:main_ardl}), is based on the `Energy Distance', with a $W=7$ day reference window \citep{szekely2007measuring, szekely2013energy}. The energy distance is a non-parametric statistic that measures the distance between two distributions based on the Euclidean distances ($\| \cdot \|_2$) between their samples. It is defined as:
\[
N_t^{\text{(energy)}} = 2\mathbb{E}[\|y_t - y_{\text{ref}}\|] - \mathbb{E}[\|y_t - y_t'\|] - \mathbb{E}[\|y_{\text{ref}} - y_{\text{ref}}'\|]
\]
where $y_t, y_t' \sim Y_t$ and $y_{\text{ref}}, y_{\text{ref}}' \sim Y_{\text{ref}}$ are drawn independently.

As a robustness check (Table \ref{tab:mmd2_ardl}), we compute an alternative measure using the Squared Maximum Mean Discrepancy (MMD$^2$) \citep{gretton2012kernel}. This measure uses a $W=30$ day reference window and is based on a kernel two-sample test. We use the Radial Basis Function (RBF) kernel, $k(x, y) = \exp(-\gamma \|x - y\|_2^2)$. The kernel's bandwidth parameter $\gamma$ is not fixed, but rather set non-parametrically for each day's comparison using the "median heuristic" (where $\gamma$ is a function of the median pairwise distance in the combined sample $Y_t \cup Y_{\text{ref}}$).

\subsection{Additional robustness results}
\label{app:robustness}

This appendix section provides supplementary evidence to bolster the main findings presented in Section~\ref{sec:results}. We demonstrate that our core results are robust to alternative model specifications, resistant to threats to inference, and specific to the theoretical mechanism under investigation. Summary statistics are presented in Appendix Table~\ref{tab:summstats}.

First, we confirm that our findings are not an artifact of our baseline ARDL($p=7, q=3$) specification. Appendix table~\ref{tab:appendix_spec_grid} details the results for alternative distributed lag lengths. 
While a model with only one lag ($q=1$) is insufficient to capture the full dynamic response, specifications allowing for a longer response window ($q=3$ and $q=7$) both yield strong, highly significant cumulative effects under HAC inference. 
This reinforces our interpretation that the strategic response is not a single-day event but rather one that unfolds and accumulates over several days.

A primary challenge to our diversionary perspective is the threat of reverse causality or spurious correlation from pre-existing trends. Our timing and placebo tests squarely address these concerns. 
Table~\ref{tab:appendix_leads} reports the results of our timing placebo test, which includes future (lead) values of the exposure variable. The cumulative sum of these leads is small and statistically indistinguishable from zero ($p = 0.553$), providing strong evidence against reverse causality. 
This null finding is further corroborated by the formal pre-trend diagnostics in Table \ref{tab:appendix_pretrend}. A joint HAC Wald test ($p = 0.686$) for the pre-shock horizons ($h \in [-5,-1]$) in our local projection model shows no evidence of a pre-existing trend. Table~\ref{tab:appendix_lp_windows} complements this by showing that, while the cumulative pre-trend window ($-3$ to $-1$) is null, the post-shock windows (e.g., $0$ to $7$) are positive and highly significant.

Finally, we validate our findings by testing their specificity and robustness to measurement choices. To refute the ``always-on circus'' perspective, we must demonstrate that the response is specific to the politically damaging Epstein coverage, not just any salient media topic. Table~\ref{tab:falsification_tv} presents this falsification test. When we replace our Epstein exposure variable with placebo keywords (``Taylor Swift'' or ``NCAA basketball''). This suggests a targeted, strategic response, but does not fully rule out the possibility of some unobserved confounder.

The finding is also not dependent on our specific operationalization of novelty or exposure. Table~\ref{tab:mmd2_ardl} shows that our main result holds when using an entirely different distributional distance metric (Squared Maximum Mean Discrepancy) to measure novelty. Table~\ref{tab:appendix_exposure} provides crucial nuance regarding the exposure source. It confirms the main finding using both Fox News density and raw mention counts. More importantly, it shows the effect is strong and significant for Google Trends ($p < 0.001$ under HAC inference) but is null for hostile media outlets (MSNBC and CNN). This corroborates the interpretation in the main text: the diversionary strategy appears to be triggered not by enemy criticism, but by signals that the damaging narrative has penetrated friendly media ecosystems or captured the attention of the broader public.

Taken together, these supplementary results provide a robust evidentiary foundation for our central claim, demonstrating that while the aforementioned caveats about causal inference still apply, the observed relationship is temporally precise, specific to the threat, and not an artifact of model specification or measurement.

\begin{table}
\centering
\small
\caption{Summary statistics}
\label{tab:summstats}
\begin{tabular}{lrrrrrrrrrr}
\toprule
variable & count & mean & std & min & p25 & median & p75 & max & missing & missing pct \\
\midrule
Energy ($N_t$) & 359 & 0.028 & 0.041 & 0 & 0.008 & 0.017 & 0.033 & 0.368 & 21 & 0.055 \\
n posts & 380 & 13.845 & 11.237 & 1 & 7 & 11 & 19 & 131 & 0 & 0 \\
 &  &  &  &  &  &  & & & 0 &  \\
\multicolumn{11}{l}{Fox News TV transcripts:} \\
hits & 386 & 2.907 & 7.395 & 0 & 0 & 0 & 2 & 59 & 0 & 0 \\
words & 386 & 84398.005 & 10722.499 & 0 & 81105 & 85373 & 89638 & 99231 & 0 & 0 \\
shows & 386 & 23.671 & 3.211 & 0 & 24 & 24 & 24 & 48 & 0 & 0 \\
shows with hits & 386 & 1.220 & 2.244 & 0 & 0 & 0 & 2 & 14 & 0 & 0 \\
density per 1000 & 386 & 0.034 & 0.087 & 0 & 0 & 0 & 0.024 & 0.671 & 0 & 0 \\
\bottomrule
\end{tabular}
\end{table}
\begin{table}
\centering

\caption{ARDL specification grid for the novelty outcome.}
\label{tab:appendix_spec_grid}
\begin{tabular}{lrrrlrr}
\toprule
Spec & $\widehat{\beta}_{\text{sum}}$ & HAC s.e. & $p_{\text{HAC}}$ & Sample & $n$ & $R^2$ \\
\midrule
$q=1$ & 0.104 & 0.061 & 0.090 & 2024-10-08--2025-10-17 & 256 & 0.198 \\
$q=3$ & 0.285 & 0.077 & <0.001 & 2024-10-08--2025-10-17 & 256 & 0.226 \\
$q=7$ & 0.481 & 0.124 & <0.001 & 2024-10-08--2025-10-17 & 256 & 0.251 \\
\bottomrule
\end{tabular}
\caption*{Columns report the short-run exposure effect alongside HAC inference, sample window, and fit statistics.}
\end{table}

\begin{table}
\centering

\caption{Timing placebo tests using leads of the exposure variable.}
\label{tab:appendix_leads}
\begin{tabular}{lrrrlrr}
\toprule
Spec & $\widehat{\delta}_{\text{sum}}$ & HAC s.e. & $p_{\text{HAC}}$ & Sample & $n$ & $R^2$ \\
\midrule
Fox density ($q=3$) & -0.042 & 0.071 & 0.553 & 2024-10-08--2025-10-16 & 255 & 0.234 \\
MMD$^2$ novelty ($q=3$) & -0.066 & 0.059 & 0.269 & 2024-10-08--2025-10-16 & 255 & 0.244 \\
\bottomrule
\end{tabular}
\caption*{Estimates report the sum of lead coefficients along with HAC inference and model diagnostics.}
\end{table}

\begin{table}
\centering

\caption{Local-projection cumulative windows for the novelty response.}
\label{tab:appendix_lp_windows}
\begin{tabular}{lrrrrlr}
\toprule
Window & Estimate & HAC s.e. & $t$ & $p_{\text{HAC}}$ & Uniform band excl. 0? & $n$ \\
\midrule
-3–-1 & 0.105 & 0.115 & 0.909 & 0.363 & -- & 266 \\
0–1 & 0.181 & 0.094 & 1.925 & 0.054 & -- & 246 \\
0–3 & 0.366 & 0.123 & 2.981 & 0.003 & -- & 228 \\
0–7 & 0.783 & 0.185 & 4.231 & <0.001 & -- & 197 \\
0–14 & 0.326 & 0.160 & 2.042 & 0.041 & -- & 160 \\
\bottomrule
\end{tabular}
\caption*{Rows list the estimated impulse responses over selected horizons together with HAC inference and whether the 95\% uniform confidence band excludes zero.}
\end{table}

\begin{table}
\centering

\caption{Falsification checks replacing Epstein exposure with alternative transcript keywords.}
\label{tab:falsification_tv}
\begin{tabular}{lrrrr}
\toprule
Keyword & $\widehat{\beta}_{\text{sum}}$ & HAC s.e. & $p_{\text{HAC}}$ & $n$ \\
\midrule
Taylor Swift & -0.136 & 0.082 & 0.098 & 256 \\
NCAA basketball & 0.026 & 0.055 & 0.635 & 256 \\
\bottomrule
\end{tabular}
\caption*{Effects are expected to be null if the novelty response is specific to the Epstein coverage spike.}
\end{table}

\begin{table}
\centering

\caption{Alternative novelty measure (MMD$^2$) using Fox exposure.}
\label{tab:mmd2_ardl}
\begin{tabular}{lrrrrr}
\toprule
Spec & $\widehat{\beta}_{\text{sum}}$ & HAC s.e. & $p_{\text{HAC}}$ & $n$ & $R^2$ \\
\midrule
$q=1$ & 0.058 & 0.056 & 0.296 & 256 & 0.201 \\
$q=3$ & 0.238 & 0.067 & <0.001 & 256 & 0.235 \\
$q=7$ & 0.475 & 0.111 & <0.001 & 256 & 0.265 \\
\bottomrule
\end{tabular}
\caption*{Entries mirror Table~\ref{tab:main_ardl}.}
\end{table}

\begin{table}
\centering

\caption{Robustness of the short-run novelty response across alternative exposure measures.}
\label{tab:appendix_exposure}
\begin{tabular}{lrrrlrr}
\toprule
Exposure & $\widehat{\beta}_{\text{sum}}$ & HAC s.e. & $p_{\text{HAC}}$ & Sample & $n$ & $R^2$ \\
\midrule
Fox News density & 0.285 & 0.077 & <0.001 & 2024-10-08--2025-10-17 & 256 & 0.226 \\
Fox News mentions & 0.277 & 0.076 & <0.001 & 2024-10-08--2025-10-17 & 256 & 0.224 \\
MSNBC density & 0.100 & 0.064 & 0.117 & 2024-10-08--2025-10-17 & 256 & 0.183 \\
CNN density & 0.109 & 0.066 & 0.100 & 2024-10-08--2025-10-17 & 256 & 0.206 \\
Cable mean (Fox+MSNBC+CNN) & 0.128 & 0.064 & 0.045 & 2024-10-08--2025-10-17 & 256 & 0.198 \\
Google Trends: Epstein & 0.320 & 0.085 & <0.001 & 2024-11-04--2025-10-17 & 229 & 0.310 \\
\bottomrule
\end{tabular}
\caption*{All specifications use the novelty outcome $N_t$ with $q=3$ exposure lags.}
\end{table}

\begin{table}
\centering

\caption{Joint pre-trend diagnostics accompanying the local-projection event study.}
\label{tab:appendix_pretrend}
\begin{tabular}{lrrrl}
\toprule
Test & Statistic & $\text{df}$ & p-value & Notes \\
\midrule
Wald $H_0$: no pre-trend & 3.090 & 5 & 0.686 &  \\
\bottomrule
\end{tabular}
\caption*{Rows report Wald tests over $h \in [-5,-1]$.}
\end{table}

\begin{table}[h!]\centering
\caption{Novelty diagnostics and posting intensity}
\label{tab:post_intensity}
\begin{threeparttable}
\begin{tabular}{l r}
\hline
Days (total) & 384 \\
Missing $N_t$ (days; share) & 25; 0.065 \\
Low-sample days & 25 \\
Zero-post days & 4 \\
Posts/day (median [p10, p90]) & 11 [4, 28] \\
Corr($N_t$, volume) & -0.054 \\
Corr($N_t$, posts) & -0.054 \\
\hline
\end{tabular}
\begin{tablenotes}[flushleft]\footnotesize
\item $N_t$ computed only when day $t$ has $\geq 3$ posts and the trailing window contributes $\geq 10$ posts; otherwise $N_t$ is missing and the day is flagged ``low-sample.'' ``Zero-post days'' are the subset with no posts. ``Posts/day” summarizes the cross-day distribution. Correlations are Pearson between standardized $N_t$ and same-day volume/post count on non-missing days.
\end{tablenotes}
\end{threeparttable}
\end{table}

\begin{table}[h!]\centering
\caption{Top novelty days (z-scored)}
\label{tab:top_novelty}
\begin{threeparttable}
\begin{tabular}{l r r}
\hline
Date & $N_{t,z}$ & Posts \\
\hline
2024-11-06 & 8.31 & 13 \\
2025-07-17 & 7.37 & 15 \\
2025-08-09 & 5.87 & 27 \\
2025-01-25 & 4.26 & 26 \\
2025-06-08 & 3.98 & 3 \\
\hline
\end{tabular}
\begin{tablenotes}[flushleft]\footnotesize
\item $N_{t,z}$ is $N_t$ standardized over the sample; ``Posts'' is the number used to compute $N_t$ on that date. High-$N_{t,z}$ dates are substantively unusual rather than volume spikes.
\end{tablenotes}
\end{threeparttable}
\end{table}

\begin{landscape}
\begin{figure}[!htbp]
  \centering
  \includegraphics[width=0.99\linewidth]{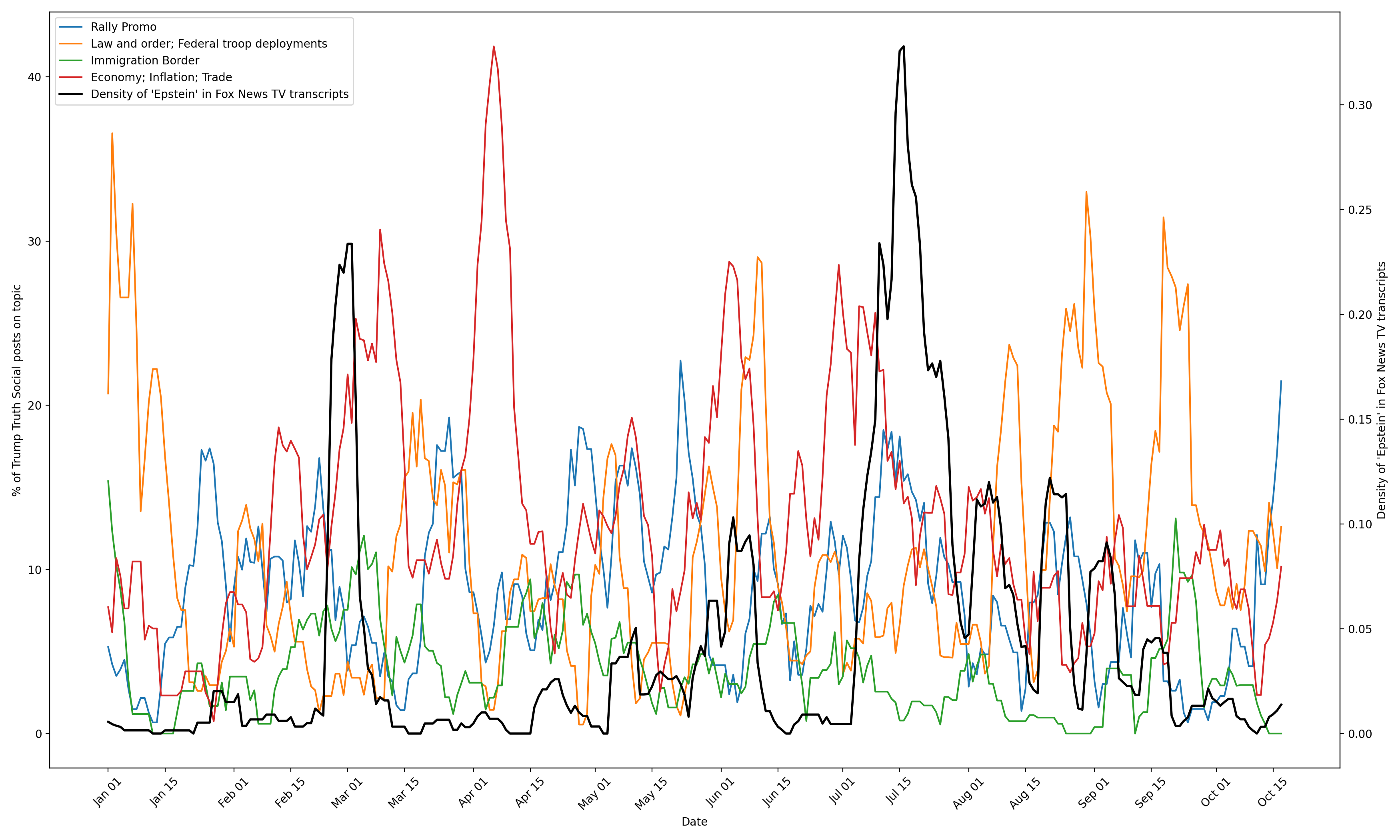}
  \caption{Density of ``Epstein'' on Fox News TV and Selected Categories of Trump Truth Social Posts.}
  \label{fig:truth_bucket_epstein_daily}
\end{figure}
\end{landscape}

\begin{table}[!htbp]
\centering
\small
\caption{Key dates, events, and sources}
\label{tab:key_dates_sources}
\begin{tabularx}{\textwidth}{l X X}
\toprule
\textbf{Date} & \textbf{Event} & \textbf{Source (title, outlet, date)} \\
\midrule
2024-01-04 & First tranche of Epstein-related court records unsealed (Maxwell defamation case). & \emph{Court documents naming Jeffrey Epstein's associates unsealed}. \href{https://abcnews.go.com/US/court-documents-naming-jeffrey-epsteins-associates-unsealed-wednesday/story?id=106047160}{ABC News}, 2024-01-04. \\
\addlinespace
2025-01-23 & Executive order directing declassification of JFK/RFK/MLK records. & \emph{Executive Order 14176-Declassification of Records Concerning the Assassinations of President John F. Kennedy, Senator Robert F. Kennedy, and the Reverend Dr. Martin Luther King, Jr.} \href{https://www.presidency.ucsb.edu/documents/executive-order-14176-declassification-records-concerning-the-assassinations-president}{The American Presidency Project (UCSB)}, 2025-01-23. \\
\addlinespace
2025-03-01 & Executive order declaring English the official U.S. language. &
\emph{Designating English as the Official Language of the United States}. \href{https://www.whitehouse.gov/presidential-actions/2025/03/designating-english-as-the-official-language-of-the-united-states/}{The White House}, 2025-03-01. \\
\addlinespace
2025-03-19 & NARA posts implementation details responding to EO 14176. & \emph{The President John F. Kennedy Assassination Records Collection (Response to Executive Order 14176)}. \href{https://www.archives.gov/research/jfk}{U.S. National Archives (NARA)}, 2025-03-19. \\

\addlinespace
2025-06-04 & Trump reinstates U.S. travel ban, barring citizens of 12 countries. &
\emph{Trump reinstates US travel ban, bars citizens of 12 countries}. \href{https://www.reuters.com/world/americas/trump-signs-proclamation-banning-travel-12-countries-cbs-news-reports-2025-06-04/}{Reuters}, 2025-06-04. \\

\addlinespace
2025-06-07 & Los Angeles deployment: announces deployment/federalization of National Guard over state objection. &
\emph{Statement from the White House}. \href{https://www.whitehouse.gov/briefings-statements/2025/06/statement-from-the-white-house-d320/}{The White House}, 2025-06-07; \emph{With troops and protests, Trump’s feud with California escalates}. \href{https://calmatters.org/politics/2025/06/los-angeles-national-guard-trump/}{CalMatters}, 2025-06-09. \\
\addlinespace
2025-07-16 & Trump dismisses ``Epstein files'' pressure as a hoax; coverage notes intra-party backlash. & \emph{Trump calls Epstein conspiracy a ‘hoax’ and turns on Maga ‘weaklings’}. \href{https://www.theguardian.com/us-news/2025/jul/16/donald-trump-dismisses-inquiry-into-jeffrey-epstein-as-boring}{The Guardian}, 2025-07-16. \\
\addlinespace

2025-08-11 & Presidential memorandum mobilizing D.C. National Guard (``Restoring Law and Order in the District of Columbia''). & \emph{Restoring Law and Order in the District of Columbia}. \href{https://www.whitehouse.gov/presidential-actions/2025/08/restoring-law-and-order-in-the-district-of-columbia/}{The White House}, 2025-08-11. \\
\addlinespace
2025-08-18 & DOJ informs House Oversight it will begin producing Epstein-related records. & \emph{Chairman Comer: DOJ Complying with Epstein Records Subpoena}. \href{https://oversight.house.gov/release/chairman-comer-doj-complying-with-epstein-records-subpoena/}{U.S. House Committee on Oversight and Accountability}, 2025-08-18. \\
\addlinespace
2025-09-02 & House Oversight releases 33{,}295 DOJ-provided pages (``Epstein records''). & \emph{Oversight Committee Releases Epstein Records Provided by the Department of Justice}. \href{https://oversight.house.gov/release/oversight-committee-releases-epstein-records-provided-by-the-department-of-justice/}{U.S. House Committee on Oversight and Accountability}, 2025-09-02. \\
\addlinespace
2025-09-08 & House Oversight releases records provided by the Epstein estate. & \emph{Oversight Committee Releases Records Provided by the Epstein Estate, Chairman Comer Provides Statement}. \href{https://oversight.house.gov/release/oversight-committee-releases-records-provided-by-the-epstein-estate-chairman-comer-provides-statement/}{U.S. House Committee on Oversight and Accountability}, 2025-09-08. \\

\addlinespace
2025-09-30 & Portland deployment: announces federal deployment/resources for public order. &
\emph{President Trump Deploys Federal Resources to Crush Violent Radical Left Terrorism in Portland}. \href{https://www.whitehouse.gov/articles/2025/09/president-trump-deploys-federal-resources-to-crush-violent-radical-left-terrorism-in-portland/}{The White House}, 2025-09-30. \\

\addlinespace
2025-10-04 & Chicago deployment: memorandum invoking federal security posture; National Guard federalization moves. &
\emph{Department of War Security for the Protection of Federal Personnel and Property in Illinois}. \href{https://www.whitehouse.gov/presidential-actions/2025/10/department-of-war-security-for-the-protection-of-federal-personnel-and-property-in-illinois/}{The White House}, 2025-10-04; \emph{Hundreds of Illinois National Guard troops to be called for federal service, memo says}. \href{https://www.cbsnews.com/chicago/news/300-illinois-national-guard-troops-federal-service/}{CBS Chicago}, 2025-10-05. \\

\addlinespace
2025-10-17 & Additional release: transcript from Alex Acosta interview published. & \emph{Oversight Committee Releases Acosta Transcript}. \href{https://oversight.house.gov/release/oversight-committee-releases-acosta-transcript/}{U.S. House Committee on Oversight and Accountability}, 2025-10-17. \\

\bottomrule
\end{tabularx}
\end{table}

\begin{table}
\caption{Example Media Claims of Diversionary tactics}
\label{tab:diversions}
\setlength{\tabcolsep}{6pt}
\renewcommand{\arraystretch}{1.2}
\begin{tabularx}{\textwidth}{
  >{\raggedright\arraybackslash}X
  >{\raggedright\arraybackslash}X
  >{\raggedright\arraybackslash}X
}
\toprule
\textbf{Category} & \textbf{Name} & \textbf{Sources} \\
\midrule

Deep-state/Scapegoating & Obama ``treason'' claim + AI arrest video &
\footnotesize
\href{https://www.reuters.com/world/us/trump-accuses-obama-treason-escalating-attacks-over-2016-russia-probe-2025-07-23/}{Reuters - ``Trump accuses Obama of treason'' (Jul 23, 2025)};
\href{https://www.washingtonpost.com/politics/2025/07/21/trump-epstein-distractions/}{Washington Post - ``As MAGA world focuses on Epstein…'' (Jul 25, 2025)};
\href{https://www.theguardian.com/us-news/2025/jul/25/donald-trump-epstein-distraction}{The Guardian - ``Distraction machine…'' (Jul 25, 2025)} \\

\addlinespace

Culture-war/Sports & Threat to block Commanders stadium unless name reverts to ``Redskins'' &
\footnotesize
\href{https://www.washingtonpost.com/opinions/2025/07/22/trump-epstein-commanders-redskins-names/}{Washington Post (Opinion) - ``Sleight of Trump: Forget Epstein! Get rid of `Commanders’!'' (Jul 22, 2025)};
\href{https://www.washingtonpost.com/politics/2025/07/21/trump-epstein-distractions/}{Washington Post - ``As MAGA world focuses on Epstein…'' (Jul 25, 2025)};
\href{https://www.theguardian.com/us-news/2025/jul/25/donald-trump-epstein-distraction}{The Guardian - ``Distraction machine…'' (Jul 25, 2025)} \\

\addlinespace

Selective transparency/Document dump & Release of MLK Jr. assassination files &
\footnotesize
\href{https://www.washingtonpost.com/politics/2025/07/21/mlk-jr-files-assassination-fbi-trump/}{Washington Post - ``Releases thousands of MLK Jr. files'' (Jul 21, 2025)};
\href{https://www.theguardian.com/us-news/2025/jul/25/donald-trump-epstein-distraction}{The Guardian - ``Distraction machine…'' (Jul 25, 2025)} \\
\addlinespace

Economic theatre/Institutional provocation & Trip to the Federal Reserve; public tussle with Chair Powell &
\footnotesize
\href{https://www.reuters.com/world/us/trumps-distraction-methods-fall-flat-against-epstein-uproar-2025-07-26/}{Reuters - ``Distraction methods fall flat…'' (Jul 26, 2025)} \\

\addlinespace

Media attack/Litigation & \$10B defamation suit against Wall Street Journal over Epstein birthday-letter report &
\footnotesize
\href{https://www.reuters.com/legal/litigation/trumps-wall-street-journal-suit-over-epstein-story-faces-timing-hurdle-2025-07-22/}{Reuters - ``WSJ suit faces timing hurdle'' (Jul 22, 2025)};
\href{https://apnews.com/article/trump-jeffrey-epstein-grand-jury-justice-department-ece8a837f9bd179771f801a765e242e4}{AP - ``Trump sues WSJ and Murdoch'' (Jul 18, 2025)} \\

\addlinespace

Celebrity/pop-culture outrage & Floats revoking Rosie O'Donnell's citizenship (Truth Social) &
\footnotesize
\href{https://www.theguardian.com/us-news/2025/jul/25/donald-trump-epstein-distraction}{The Guardian - ``Distraction machine…'' (Jul 25, 2025)} \\

\addlinespace

Flood-the-zone posting/Memes & Rapid-fire viral clips and odd posts to drown out Epstein coverage &
\footnotesize
\href{https://www.washingtonpost.com/politics/2025/07/21/trump-epstein-distractions/}{Washington Post - ``As MAGA world focuses on Epstein…'' (Jul 25, 2025)};
\href{https://www.theguardian.com/us-news/2025/jul/25/donald-trump-epstein-distraction}{The Guardian - ``Distraction machine…'' (Jul 25, 2025)} \\

\addlinespace

Press access management & Bans Wall Street Journal from Scotland trip amid Epstein furor &
\footnotesize
\href{https://www.washingtonpost.com/the-seven/2025/07/22/what-to-know-for-july-22/}{Washington Post - ``The 7: Trump's Epstein distractions…'' (Jul 22, 2025)} \\

\addlinespace

Narrative reframing & Labels Epstein issue a ``hoax''; rebukes supporters as ``weaklings'' &
\footnotesize
\href{https://www.theguardian.com/us-news/2025/jul/16/donald-trump-dismisses-inquiry-into-jeffrey-epstein-as-boring}{The Guardian - ``Calls Epstein conspiracy a `hoax’…'' (Jul 16, 2025)};
\href{https://apnews.com/article/47370e63060d79af64b0b8ac70b970a4}{AP - ``Trump slams his own supporters as `weaklings'…'' (Jul 2025)} \\

\bottomrule
\end{tabularx}
\end{table}

\end{document}